\documentstyle[12pt]{article}

\begin{document} 

\begin{center}  {\Large {\bf Relating Meson and Baryon Fragmentation Functions by Shower-Parton Recombination}}
\vskip .75cm
 {\bf Rudolph C. Hwa$^1$ and C.\ B.\ Yang$^{1,2}$}
\vskip.5cm
{$^1$Institute of Theoretical Science and Department of
Physics\\ University of Oregon, Eugene, OR 97403-5203, USA\\
\bigskip
$^2$Institute of Particle Physics, Hua-Zhong Normal
University, Wuhan 430079, P.\ R.\ China}
\end{center}

\vskip.5cm
\begin{abstract} 
We relate the fragmentation functions of partons into mesons and baryons in the framework of recombination of shower partons. The results are in reasonable agreement with the data. The implication is that the meson and baryon fragmentation functions are not independent when hadronization of the shower partons are taken into account. The conclusion therefore closes a conceptual gap in the system of fragmentation functions whose $Q^2$ evolution has been more extensively studied than their interrelationship.
\end{abstract}

\section{Introduction}

Hadron production at intermediate $p_T$ in heavy-ion collisions (HIC) has been successfully described in the recombination model (RM) \cite{hy,gkl,fmnb}.  In Ref.\  \cite{hy} the hadronization process is shown to be dominated by the recombination of thermal and shower partons when $p_T$ is in the range $3 < p_T < 8$ GeV/c.  The distributions of those shower partons are derived by fitting the fragmentation functions (FFs) in the RM \cite{hy2}.  The shower partons considered are not to be identified with those due to gluon radiation at very high virtuality that are calculable in pQCD, since the ones we consider are defined by their recombination within a jet to form a hadron at the hadronization scale, and are not calculable in pQCD.  In this paper we revisit the problem of fragmentation in the framework of recombination, and calculate the FFs for proton and $\Lambda$ using the shower parton distributions (SPDs)
 that have been determined in Ref.\ \cite{hy2} from fitting the FFs for mesons.
 
 The determination of FF for proton in a jet that can agree with data would be a significant achievement, since it is the first time a  FF is obtained by calculation, instead of by fitting.  It should be stated from the outset that we are not concerned with $Q^2$ evolution here, since that part of the problem is well handled by pQCD \cite{bkk}-\cite{sk2}.  All evolutions have to start from somewhere, and the starting distributions have to be determined by fitting the data.  In Refs. \cite{kkp,kkp2} the FFs have been parameterized for a variety of initiating partons and final hadrons, and have all been determined independently apart from trivial identities.  To our knowledge no one has attempted to relate one FF to another.  Our attempt to relate $D^{\pi}(z)$ and $D^{p}(z)$, FFs for $\pi$ and $p$, respectively, is feasible because our emphasis is on the hadronization of the shower partons.  Thus a successful connection among the FFs is an affirmation of the soundness of the recombination formalism that we employ, and establishes an important link among the multitude of FFs.

\section{Gluon Fragmentation into Proton}

We first recall how we have treated the fragmentation of gluon into pion in the framework of recombination.  The basic equation is \cite{hy2}
\begin{eqnarray}
x D^{\pi}_G (x) =  \int {dx_1 \over x_1} {dx_2 \over x_2} \left\{G(x_1), G(x_2) \right\} R_{\pi} (x_1, x_2, x)\ ,
\label{1}
\end{eqnarray}
where $D^{\pi}_G (x)$ is the FF of gluon into pion, and $R_{\pi} (x_1, x_2, x)$ is the recombination function (RF) for $q\bar{q} \to \pi$ that is given in Ref.\ \cite{hy,hy4}.  $G(x_i)$ is the SPD of $q$ or $\bar{q}$ in a gluon jet, and the symmetrization symbol in Eq.\ (\ref{1}) means
\begin{eqnarray}
 \left\{G(x_1), G(x_2) \right\} = {1 \over 2}\left[ G(x_1)G\left(x_2 \over 1 - x_1 \right)+G(x_2)G\left(x_1 \over 1 - x_2 \right)\right] \ .
 \label{2}
\end{eqnarray}
Since $D^{\pi}_G (x)$ is known from the parametrization by  BKK \cite{bkk} determined by fitting various data, we can solve for $G (z)$ in Eq.\ (\ref{1}), not analytically, but by fitting the FF at a chosen $Q^2$.  The form for $G (z)$ used is 
\begin{eqnarray}
G (z) = A z^a (1-z)^b (1+cz^d) \ .
 \label{3}
\end{eqnarray}
For $Q^2 = 100$ GeV$^2$ the values of the parameters are $A = 0.811$, $a = - 0.056$, $b = 2.547$, $c = - 0.176$, and $d = 1.2$ \cite{hy2}.  The evolution of these parameters for other values of $Q^2$ has been considered elsewhere \cite{ty}.  It should be emphasized that $G(z)$ is defined by Eq.\ (\ref{1}) in the context of parton recombination, and is the distribution at the hadronization scale of a light quark (or antiquark) in a gluon-initiated jet.  The $q$ and $\bar{q}$ at $x_1$ and $x_2$ are to recombine to form a pion at $x$, which is constrained by $R_{\pi} (x_1, x_2, x)$ to be $x = x_1 + x_2$.

We now advance the point of view that if Eq.\ (\ref{1}) makes sense in describing FF in terms of the recombination of shower partons, then the same formalism should make possible a calculation of the FF of gluon into proton.  Indeed, we need only generalize Eqs.\ (\ref{1}) and (\ref{2}) in the self-evident way
\begin{eqnarray}
x D^{'p}_G (x) =  \int {dx_1 \over x_1} {dx_2 \over x_2}{dx_3 \over x_3} \left\{G(x_1), G(x_2), G(x_3) \right\} R_p (x_1, x_2, x_3, x)\ ,
 \label{4}
\end{eqnarray}
where
\begin{eqnarray}
 \left\{G(x_1), G(x_2), G(x_3) \right\}  = {1 \over 6} \left[ G(x_1) G\left({x_2 \over 1 - x_1}\right) G\left({x_3 \over 1 - x_1-x_2}\right) + \dots \right]\ ,
 \label{5}
\end{eqnarray}
there being six terms in the permutation of $x_1, x_2$ and $x_3$.  $D^{'p}_G (x)$ is an initial form of the FF of gluon into proton, on which modification will be made in the following.  $R_p$ is the RF for the proton \cite{hy}
\begin{eqnarray}
 R_p (x_1, x_2, x_3, x) = g_{st} g_p \left( {x_1,x_2 \over x^2}\right)^{\alpha + 1}     \left( {x_3 \over x}\right)^{\beta + 1}  \delta \left({x_1+ x_2+ x_3  \over x} -1\right)  \ ,
 \label{6}
\end{eqnarray}
where $g_{st} = 1/6$ and $g_p = [B(\alpha + 1, \alpha + \beta +2)B(\alpha + 1,  \beta +1)]^{-1}$, $B(\alpha,\beta)$ being the Euler-beta function.  The parameters $\alpha$ and $\beta$ are \cite{hy5}
\begin{eqnarray}
\alpha = 1.75, \qquad \beta = 1.05 \ .
 \label{7}
\end{eqnarray}

We do not regard $x D^{'p}_G (x)$ as the final form for the FF because we have not so far taken into account the momentum carried by the antiproton that must accompany the production of a proton in a gluon jet.  To treat the pair production of $p \bar{p}$, we have to consider the correlation between the dibaryon produced.  If $D^{p \bar{p}}_G(x)$ denotes the $p \bar{p}$ FF, then upon integration over $x_2$ we should have on general grounds
\begin{eqnarray}
\int dx_2 D^{p \bar{p}}_G(x_1, x_2) = D^{p}_G(x_1) \int dx_2 D^{\bar{p}}_G(x_2)
 \label{8}
\end{eqnarray}
which does not imply factorizability of $D^{p \bar{p}}_G(x_1, x_2)$.  We shall apply the momentum constraint $x_1 + x_2 < 1$ as the only correlation between $p$ and $\bar{p}$, and write
\begin{eqnarray}
D^{p \bar{p}}_G(x_1, x_2) = {1 \over 2} \left[ D'^p_G (x_1) D'^{\bar{p}}_G \left({x_2 \over 1 - x_1}\right) + D'^p_G \left({x_1 \over 1 - x_2}\right) D'^{\bar{p}}_G \left(x_2\right) \right]\ .
\label{9}
\end{eqnarray}
Commensurate to Eq.\ (\ref{9}), we approximate $D^{\bar{p}}_G(x_2)$ in Eq.\ (\ref{8}) by $D'^{\bar{p}}_G(x_2)$.  Then $D^p_G (x_1)$ can be evaluated from 
\begin{eqnarray}
D^p_G(x_1) = {\int dx_2 \left\{ D'^p_G(x_1), D'^{\bar{p}}_G(x_2)\right\} \over \int dx_2 D'^{\bar{p}}_G(x_2)}  \ .
\label{10}
\end{eqnarray}
The result for $xD^p_G(x)$ is shown in Fig.\ 1 by the solid line.  Also shown in that figure in dashed line is the FF obtained by KKP \cite{kkp2} by fitting the data in NLO.  The agreement over 5 orders of magnitude is very good, considering the fact that no free parameters have been used in the calculation.

\section{Other Fragmentation Functions}

In this section we consider two other FFs into baryons, first $g \to \Lambda$, and then $u \to p$.  For the former Eq.\ (\ref{4}) is replaced by
\begin{eqnarray}
x D'^{\Lambda}_G (x) =  \int \left(\prod^3_{i = 1} {dx_i \over x_i} \right) \left\{G(x_1), G(x_2), G_s(x_3) \right\} R_{\Lambda} (x_1, x_2, x_3, x)\ ,
 \label{11}
\end{eqnarray}
where $G_s (x_3)$ is the SPD of a strange quark in a gluon jet, and $R_{\Lambda} (x_1, x_2, x_3, x)$ is the RF for $\Lambda$.  In Ref.\ \cite{hy2}, $G_s(z)$ is parameterized as in Eq.\ (\ref{3}), but with $A = 0.069$, $a = -0.425$, $b = 2.489$, $c = -0.5$ and $d = 1.1$.  $R_{\Lambda}$ is given in Ref. \cite{hy6}; it has the same form as Eq.\ (\ref{6}) but with $g_{st} = 1/4$, $g_{\Lambda}=g_p$, $\alpha = 1$ and $\beta =2$.  Thus $D'^{\Lambda}_G (x) $ can be calculated using Eq.\ (\ref{11}) and then modified as before to give $D^{\Lambda}_G (x) $ by taking the  momentum of $\bar\Lambda$ into account, as done in Eq.\ (\ref{10}).  The result is shown by the solid line in Fig.\ 2.  For comparison, we show in the same figure the LO and NLO distributions of $g \to \Lambda$ at $Q^2 = 100$ GeV$^2$ given in Ref.\ \cite{dsv}.  Although our result is a little lower in the mid-$x$ region, the agreement is nevertheless acceptable; in fact, it agrees better with the LO result.

Next, we consider proton production in a jet initiated by a $u$ quark.  In that case valence and sea quark distributions must be distinguished.  In Ref.\ \cite{hy2} we have used $K_{NS}$ and $L$ to denote the valence and sea quark SPDs, respectively; they are parameterized as in Eq.\ (\ref{3}) with the parameters for $K_{NS}\  (L):  A = 0.333 \ (1.881), a = 0.45 \ (0.133), b = 2.1 \ (3.384), c = 5.0 \  (-0.991),$ and $d = 0.5 \ (0.31)$.  The recombination formula is now 
\begin{eqnarray}
x D'^p_u (x) =  \int \left(\prod^3_{i = 1} {dx_i \over x_i} \right) \left\{K(x_1), L(x_2), L(x_3) \right\} R_p (x_1, x_2, x_3, x)\ ,
 \label{12}
\end{eqnarray}
where
\begin{eqnarray}
K(x_1) = K_{NS} (x_1) + L(x_1)  \ .
 \label{13}
\end{eqnarray}
For $\bar{p}$ production in the same jet valence quark does not contribute, so for $D'^{\bar{p}}_u (x)$ the sea quark distribution $L(x_1)$ replaces $K(x_1)$ in Eq.\ (\ref{12}).  The resultant $D'^P_u$ and $D'^{\bar{p}}_u$ are then substituted into 
\begin{eqnarray}
D^p_u (x_1) =  {\int  dx_2  \left\{D'^p_u (x_1),D'^{\bar{p}}_u (x_2) \right\} \over \int  dx_2 D'^{\bar{p}}_u (x_2)} \  .
 \label{14}
\end{eqnarray}
The FF thus obtained is shown by the solid line in Fig.\ 3.  The parametrization of the data by  KKP \cite{kkp2}, shown in the dashed line, is, however, unreliable because an invalid assumption, $D^p_u (z) = 2D^p_d (x)$, has been used in the analysis.  While we await an improved analysis of the data, the solid line in Fig.\ 3 may be regarded as our prediction of the FF for $u$ quark into $p$.

\section{Conclusion}

We have calculated the FFs into baryons, using the SPDs determined by fitting the FFs into mesons.  The results are in good agreement with data for $g \to p$ and $g \to \Lambda$.  For $u \to p$ there is no reliable analysis of the data, so our result is a prediction.  To be able to relate meson and baryon FFs is an attribute of our formalism that treats fragmentation as a recombination process.  Since the shower partons are defined at the hadronization scale, their distributions cannot be calculated from first principles.  Nevertheless, they have been determined by fitting FFs for mesons, inasmuch as the FFs themselves have been determined by fitting the data.  What we have achieved in this paper is to elevate the significance of the formalism that connects previously unrelated FFs.  That not only confirms the internal consistency of the formalism, but also renders additional support to the reliability of the procedure employed earlier in the study of the contribution of thermal-shower recombination to hadron production (mesons and baryons) at intermediate $p_T$ in HIC.

The experience gained through this work on baryon production builds the basis for future investigation of correlations that involve baryons produced at intermediate $p_T$ in HIC.  Data on such correlations from experiments at RHIC already exists \cite{phenix}.  Whether  a baryon is the trigger particle or an associate particle, it is clear that the recombination model is well positioned to address the problem of hadronization of various partons that arise from different sources  (hard and soft) in HIC.

\section*{Acknowledgment}

We are grateful to  W.\ Vogelsang  for helpful communication on fragmentation functions.  This work was supported  in
part,  by the U.\ S.\ Department of Energy under Grant No. DE-FG02-96ER40972 and by the Ministry of Education of China under Grant No. 03113.

\begin{center}
\section*{Figure Captions}
\end{center}

\begin{description}
\item
Fig.\ 1. Fragmentation function for gluon into proton. Solid line is our calculated result; dashed line represents the fit of the data in NLO  \cite{kkp2}.

\item
Fig.\ 2. Fragmentation function for gluon into $\Lambda$. Solid line is our calculated result. The line with circles represents the fit of the data in NLO, while the dashed line with squares is in LO  \cite{dsv}.
 
 \item
Fig.\ 3. Fragmentation function for $u$ quark into proton. Solid line is our calculated result; dashed line represents a questionable fit of the data in NLO  \cite{kkp2} (see text).

\end{description}

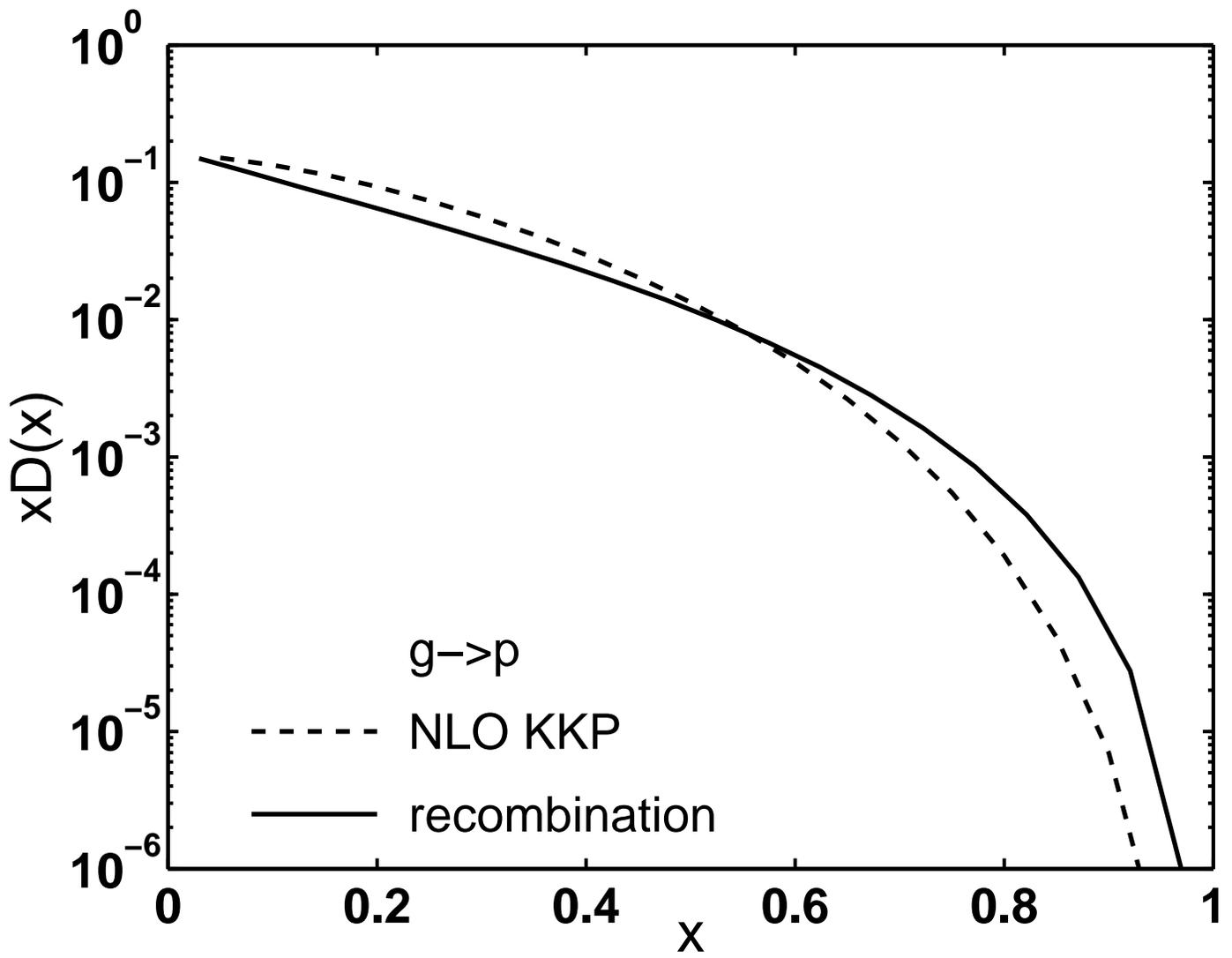
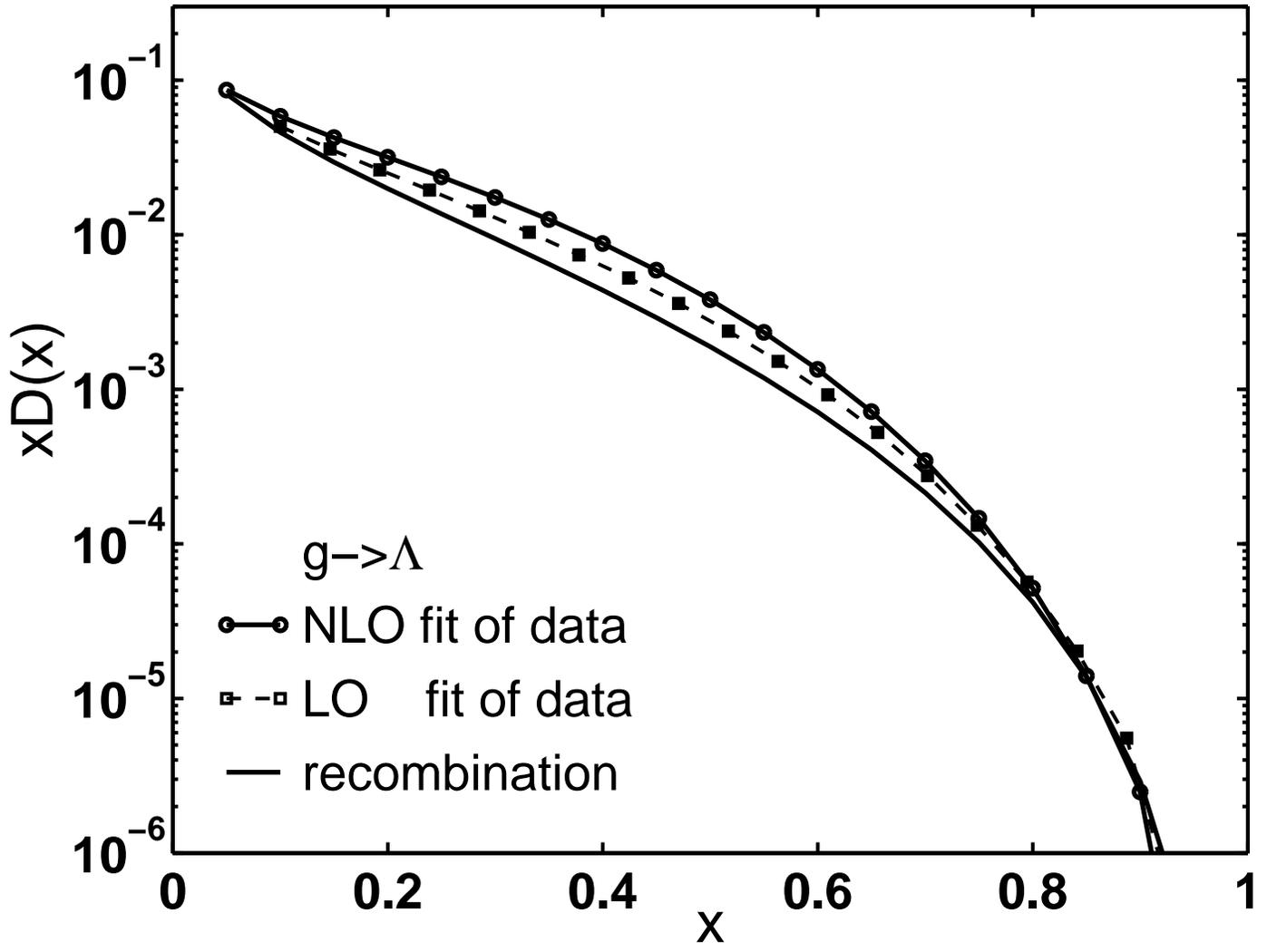
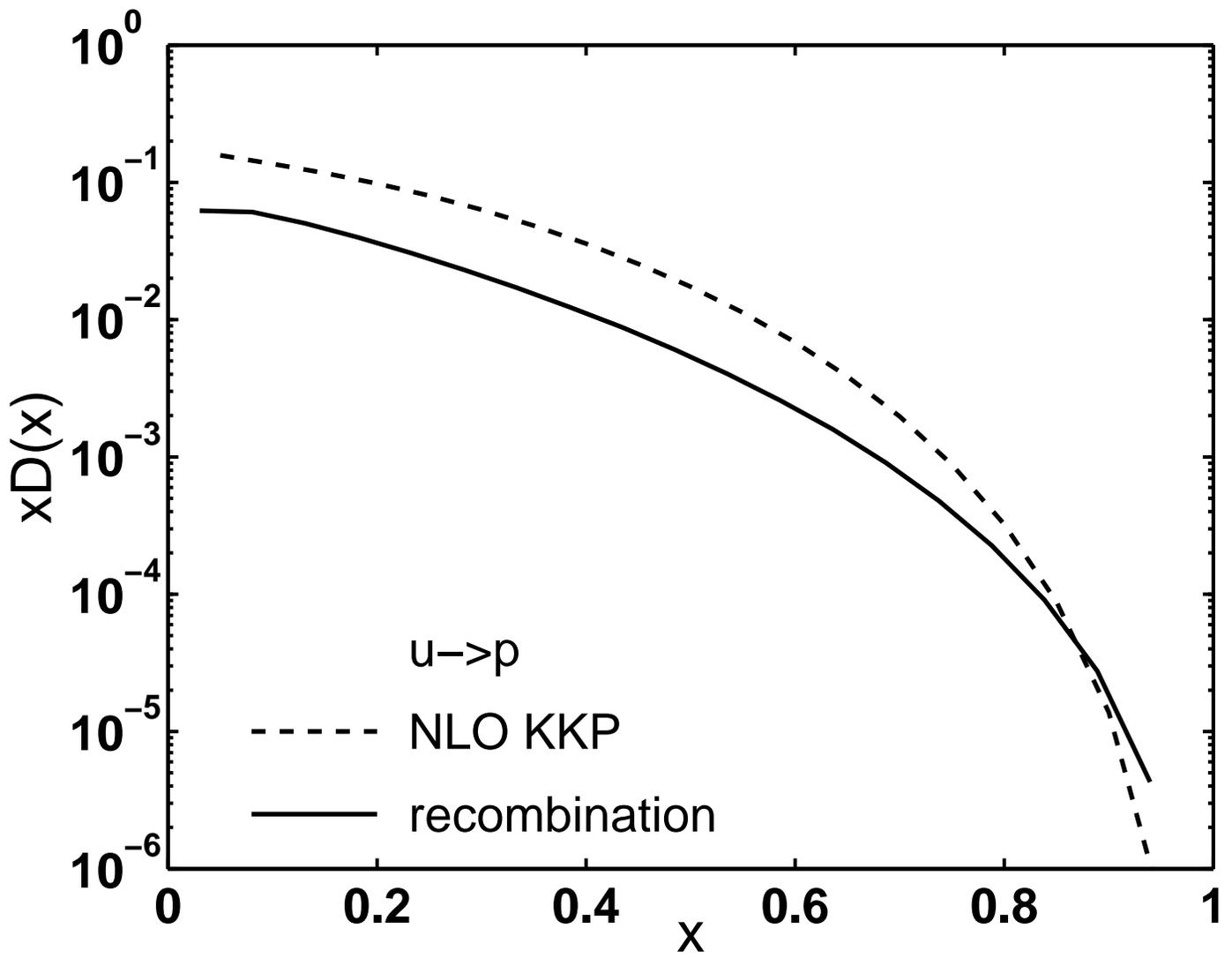
\end{document}